\documentstyle[graphicx,twoside,fleqn,espcrc2]{article}

\newcommand{\llssor}{{\it ll}\/-SSOR}

\newcommand{\beq}{\begin{equation}}
\newcommand{\eeq}{\end{equation}}        
\newcommand{\bqa}{\begin{eqnarray}}        
\newcommand{\eqa}{\end{eqnarray}}        
\newcommand{\be}{\begin{enumerate}}
\newcommand{\ee}{\end{enumerate}}        
\newcommand{\bi}{\begin{itemize}}
\newcommand{\ei}{\end{itemize}}

\newcommand{\fig}[1]{{\frenchspacing Fig.~\ref{#1}}}
\newcommand{\nn}{\nonumber}

\newcommand{\one}{{\mathbf 1}}
\newcommand{\uno}{\one}

\title{
\hfill\begin{minipage}{0pt}\scriptsize\vspace*{-1.5cm} \begin{tabbing}
\hspace*{\fill} HLRZ1998-52\\ 
\hspace*{\fill} WUP-TH 98-29 
\end{tabbing} 
\end{minipage}\\[-8pt]
A Preconditioner for Improved Fermion Actions}
     
\author{\frenchspacing N.~Eicker\thanks{Talk presented by
    N.~Eicker}\address{HLRZ, c/o Research Center J\"{u}lich, D-52425
    J\"{u}lich, Germany, and DESY/Hamburg, Germany},
  W.~Bietenholz$^{\rm{a}}$, A.~Frommer\address{Department of
    Mathematics, University of Wuppertal, D-42097 Wuppertal, Germany},
  Th.~Lippert\address{Department of Physics, University of Wuppertal,
    D-42097 Wuppertal, Germany}, B.~Medeke$^{\rm{b}}$, and
  K.~Schilling$^{\rm{a,c}}$ \nonfrenchspacing}

\begin{document}

\begin{abstract}
  SSOR preconditioning of fermion matrix inversions which is
  parallelized using a {\em locally-lexicographic} lattice
  sub-division has been shown to be very efficient for standard
  Wilson fermions. We demonstrate here the power of this method for
  the Sheikholeslami-Wohlert improved fermion action and for a
  renormalization group improved action incorporating couplings of the
  lattice fermion fields up to the diagonal in the unit hypercube.
\end{abstract}

\maketitle

\section{Introduction}

Recently, the symmetric successive over-relaxed preconditioner (SSOR)
turned out to be parallelizable by means of the {\it l}\/ocally-{\it
  l}\/exicographic ordering technique \cite{FIS96}. In this way, SSOR
preconditioning has been made applicable to the acceleration of
standard Wilson fermion inversions on high performance massively
parallel systems and it outperforms o/e preconditioning.

It appears intriguing to extend the range of \llssor-preconditioners
such as to accelerate the inversion of improved fermionic actions,
which became very popular in the recent years.

In {\em Symanzik's on-shell improvement} program \cite{SYM83}, counter
terms are added to both, lattice action and composite operators in
order to reduce ${\cal O}(a)$ artifacts which spoil results in the
instance of the Wilson fermion formulation.  In the approach of
Sheikholeslami and Wohlert (SWA) \cite{SHE85}, the Wilson action is
modified by adding a diagonal term, the so-called clover term with a new
free parameter $c_{SW}$.

{\em Perfect lattice actions} are situated on renormalized
trajectories in parameter space that intersect the critical surface
(at infinite correlation length) in a fixed point of a renormalization
group transformation.  Perfect actions are free of any cut-off
effects, but in practice they can only be constructed approximatively.
A promising approach for asymptotically free theories is the use of
classically perfect actions \cite{HAS94} to serve as an approximation
to perfect ones.  Moreover, practical applications require a
truncation of the couplings to short distances (truncated perfect
actions, TPA). In the present investigation, we consider a variant of
the hypercube fermion (HF) approximation \cite{BIE96}.

The generic form of both SWA and TPA is given by
\begin{equation}
M=D+A+B+C+E\dots
\label{eq:shape}
\end{equation}
$D$ stands for $12\times 12$ diagonal sub-blocks, $A$, $B$,\dots are
nearest-neighbor, next-to-nearest-neighbor,\dots hopping terms.  In
the following, we will show that the \llssor\ scheme applies not only
to the couplings in $A$ but also to the internal spin and colour {\em
  d.o.f.} of $D$ (SWA) as well as all the couplings of $B$, $C$, and
$E$, \dots of TPA.

\section{SWA and HF Actions}

SWA is composed of $A$ (Wilson hopping term) and $D$ (SW diagonal):
\begin{eqnarray}
  D_{SW}(x,y)= \Big[ \one +&\!\!\!\!\!\!\!\!\!\!&
  \frac{c_{SW}}{2}\kappa\sum_{\mu,\nu}
  \sigma_{\mu\nu} F_{\mu\nu}(x)\Big] \delta_{x,y}; \nn\\
  A_{SW}(x,y)= - \kappa\sum_\mu&\!\!\!  \Big[\!\!\!\!\!\!&
  (\one-\gamma_\mu)U_\mu(x)\delta_{x,y-\hat \mu}\\ +
  &\!\!\!\!\!\!\!\!& (\one+\gamma_\mu)U_{-\mu}(x) \delta_{x,y+\hat
    \mu}\Big], \nonumber
  \label{eq:clover_matrix}
\end{eqnarray}
$\kappa$ is the Wilson hopping parameter, $c_{SW}$ couples the SW
clover operator. This parameter is tuned to optimize ${\cal O}(a)$
cancellations.  The clover term $F_{\mu\nu}(x)$ consists of $12\times
12$ diagonal blocks. Its explicit structure in Dirac space is given in
Ref.~\cite{FRO98}.

As a prototype TPA we have investigated the perfect free action
constructed in Ref.~\cite{BW96} by means of block variable
renormalization group transformations for free fermions.  The
exponential decay of their couplings is fast, and therefore they can
be truncated to short range \cite{BIE96}.  We limit ourselves to
couplings up to 4-space diagonals in the unit hypercube (hypercube
fermion, HF). The gauge links are introduced in an obvious way: we
connect all the coupled sites by all possible $d!$ {\em shortest}
lattice paths in a $d$ diagonal, by multiplying the compact gauge
fields on the path links. For a given link, we average over all $d!$
paths from hyper-links $U^{(d)}_{\mu_1+\mu_2+\dots+\mu_d}(x)$ built up
recursively:
\begin{eqnarray}
\lefteqn{U^{(d)}_{\mu_1+\mu_2+\dots+\mu_d}(x)}\nn\\
=&\frac{1}{d}\Big[&
U^{(1)}_{\mu_1}(x)\,
U^{(d-1)}_{\mu_2+\mu_3+\dots+\mu_d}(x+\hat{\mu}_1) \nonumber\\
&+&U^{(1)}_{\mu_2}(x)\,
U^{(d-1)}_{\mu_1+\mu_3+\dots+\mu_d}(x+\hat{\mu}_2) \nonumber\\
&+&\dots\nonumber\\
&+&U^{(1)}_{\mu_d}(x)\,
U^{(d-1)}_{\mu_1+\mu_2+\dots+\mu_{d-1}}(x+\hat{\mu}_d)\;\;\Big].
\end{eqnarray}
Defining effective $\Gamma$'s by
\beq%
\Gamma_{\pm\mu_1\pm\mu_2\pm\dots\mu_d} = \lambda_d
+\kappa_d(\pm\gamma_{\mu_1}\pm\gamma_{\mu_2}\pm\dots\gamma_{\mu_d})\,
,\label{lamm}
\eeq%
with the HF hopping parameters $\kappa_i$ and $\lambda_i$, we arrive
at:
\begin{eqnarray}
D_{HF}(x,y)\!\!\!&\!\!\!\!\!\!=\!\!\!\!\!\!&\!\!\! \lambda_{0}\delta_{x,y}\,,\hspace{1cm} \\
  A_{HF}(x,y)\!\!\!&\!\!\!\!\!\!=\!\!\!\!\!\!&\!\!\! \sum\limits_{\mu}\Big[
\Gamma_{+\mu}U^{(1)}_\mu(x)\delta_{x,y-\hat \mu}+\nn\\[-.3cm]
&&\hspace{.6cm}\Gamma_{-\mu}U^{(1)}_{-\mu}(x)\delta_{x,y+\hat\mu}\Big]\,,\nn\\
  B_{HF}(x,y)\!\!\!&\!\!\!\!\!\!=\!\!\!\!\!\!&\!\!\!\sum\limits_{{\mu}}\sum\limits_{{\nu>\mu}}\Big[
\Gamma_{+\mu+\nu}U^{(2)}_{+\mu+\nu}(x)\delta_{x,y-\hat\mu-\hat\nu}\nn\\[-.3cm]
&&\hspace{.95cm}+\Gamma_{+\mu-\nu}U^{(2)}_{+\mu-\nu}(x)\delta_{x,y-\hat\mu+\hat\nu}\nn\\
&&\hspace{.95cm}+\Gamma_{-\mu+\nu}U^{(2)}_{-\mu+\nu}(x)\delta_{x,y+\hat\mu-\hat\nu}\nn\\
&&\hspace{.95cm}+\Gamma_{-\mu-\nu}U^{(2)}_{-\mu-\nu}(x)\delta_{x,y+\hat\mu+\hat\nu}\Big]\nn.
\label{HFMATRIX}
\end{eqnarray}
It is straightforward to write down the expressions for $C_{HF}$ and
$E_{HF}$.  Altogether 80 hyper-links contribute while 40 have to be
stored.

\section{Block SSOR Preconditioning\label{SSOR}}

The preconditioned system is modified by two matrices
$V_1$ and $V_2$,
\begin{equation}
  V_1^{-1}MV_2^{-1} \tilde{\psi} = \tilde{\phi}, \;
   \tilde{\phi} := V_1^{-1}\phi, \; \tilde{\psi} := V_2\psi.
  \label{Matrix_prec_eq}
\end{equation}
Let $M = D - L - U$ be the decomposition of $M$ into its block
diagonal part $D$, its (block) lower triangular part $-L$ and its
(block) upper triangular part $-U$.  Block SSOR preconditioning is
defined through the choice
\begin{equation}
V_1 = \left( \frac{1}{\omega}D - L \right) \left( \frac{1}{\omega}D 
\right)^{-1},
 \enspace V_2 = \frac{1}{\omega}D - U \; .
\label{DEFOMEGA}
\end{equation}
The {\em Eisenstat trick} \cite{FIS96} reduces the costs by a factor
2. It is based on the identity:
\begin{eqnarray}
&&\hspace*{-.7cm}V_1^{-1} (D - L - U) V_2^{-1}= \left( \uno\right. -\omega LD^{-1}
\left.    \right)^{-1}\times\nn\hspace{.35cm}(8)\\
\nn 
&&\hspace*{-.7cm}\left[ \uno + (\omega \!-\! 2) \left(\uno - \omega
      UD^{-1} \right)^{-1} \right]\!\!+\!\!\left( \uno - \omega UD^{-1}
  \right)^{-1}\! .
\end{eqnarray}
The preconditioned matrix-vector product,
$z = V^{-1}_1MV^{-1}_2 x$, is given by:

\vspace*{6pt}
\begin{minipage}{\columnwidth}
\begin{tabbing}
\hspace*{2ex} \= \hspace{2ex} \=  \hspace{2ex} \= \hspace{2ex} \kill
\> \sf solve $(\uno - \omega UD^{-1}) y = x$ \\
\> \sf compute $w = x + (\omega - 2) y $ \\
\> \sf solve $(\uno - \omega LD^{-1}) v = w$ \\
\> \sf compute $z = v + y$
\end{tabbing}  
\end{minipage}
\vspace*{6pt}

The {\sf ``solve''} is just a simple forward (backward) substitution
process due to the triangular structure:

\vspace*{6pt}
\begin{minipage}{\textwidth}
\begin{tabbing}
\hspace*{2ex} \= \hspace{2ex} \=  \hspace{2ex} \= \hspace{2ex} \kill
\> for $i=1$ to $N$ \\
\> \> $v_i = w_i + \sum_{j=1}^{i-1} L_{ij} s_j$  \\
\> \> $s_i = \omega D_{ii}^{-1}v_i$ 
\end{tabbing} 
\end{minipage}
\vspace*{6pt}

Options for $D$ of SWA take each block $D_{ii}$ to be of dimension 12
($D^{(12)}$), 6 ($D^{(6)}$), 3 ($D^{(3)}$) or 1 ($D^{(1)}$), as
suggested by the structure of $D$.  The blocks have to be pre-inverted
the costs depending on the block size \cite{FRO98}.

Parallelism can be achieved by {\em locally lexicographic} ordering
\cite{FIS96}.  ``Coloring'' is the decomposition of all lattice points
into mutually disjoint sets $C_1,\ldots,C_k$ (with respect to the
matrix $M$), if for any $l \in \{1,\ldots,k\}$ the property $ x \in
C_l \Rightarrow y \not \in C_l\; \mbox{ for all }\; y \in n(x) $
holds.  $n(x)$ denotes the set of sites $\ne x$ coupled to $x$.  A
suitable ordering first numbers all $x$ with color $C_1$, then all
with $C_2$ etc. Thus, each lattice point couples with lattice points
of different colors only.  The computation of $v_x$ for all $x$ of a
given color $C_l$ can be done in parallel, since terms like $\sum_{y
  \in n(x), \; y \leq_o x }$ involve only lattice points of the
preceding colors $C_1,\ldots,C_{l-1}$, with $x \leq_{o} y$ meaning
that $x$ has been numbered before $y$ with respect to the ordering
$o$.

Let the  lattice blocks be of size $n^{loc}=n^{loc}_1 \times n^{loc}_2
\times n^{loc}_3 \times n^{loc}_4$. A different color is associated
with each of the sites of the $ n^{loc}$ groups. A {\em locally
  lexicographic} ($ll$) ordering is defined to be the color
ordering, where all points of a given color are ordered after all
points with colors, which correspond to lattice positions on the local
grid that are lexicographically preceding the given color. The
parallel forward substitution reads:

\vspace*{6pt}
\begin{minipage}{\textwidth}
\begin{tabbing}
  \hspace*{2ex} \= \hspace{2ex} \= \hspace{2ex} \= \hspace{2ex} \kill
  for $C_i$, $i=1,\dots,\frac{n}{p}$, $frac{n}{p}\in {\bf N}$\\
  \> for all processors $j=1,\dots,p$\\
  \>  \> $x := $ with $C_i$ on $j$ \\
  \> \> $v_x = w_x + \sum_{y \in n(x), \; y \leq_{ll} x } L_{xy} s_y$ \\
  \> \> $s_x = \omega D_{xx}^{-1}v_x$
\end{tabbing}
\end{minipage}
\vspace*{6pt}
 
For SWA, up to 8 and for HF all 80 neighbors may be involved on the
4-d grid \cite{FIS96}.

\section{Improvement\label{RESULTS}}
The SWA has been implemented on an APE100. HF is benchmarked on a DEC
alpha workstation.  For SWA, we use a de-correlated set of 10 quenched
gauge configurations generated on a $16^4$ lattice at $\beta=6.0$ at 3
values of $c_{SW}$, 0, 1.0 and 1.769.  We have applied BiCGStab as
iterative solver.  The stopping criterion has been chosen as
$\frac{||MX-\phi||}{||X||}\le 10^{-6}$, with $X$ being the solution.
We used a local source $\phi$ and determined the optimal OR parameter
to be about $\omega=1.4$ for all block sizes and $c_{SW}$.

\begin{figure}[!htb]
{\includegraphics[width=.92\columnwidth]{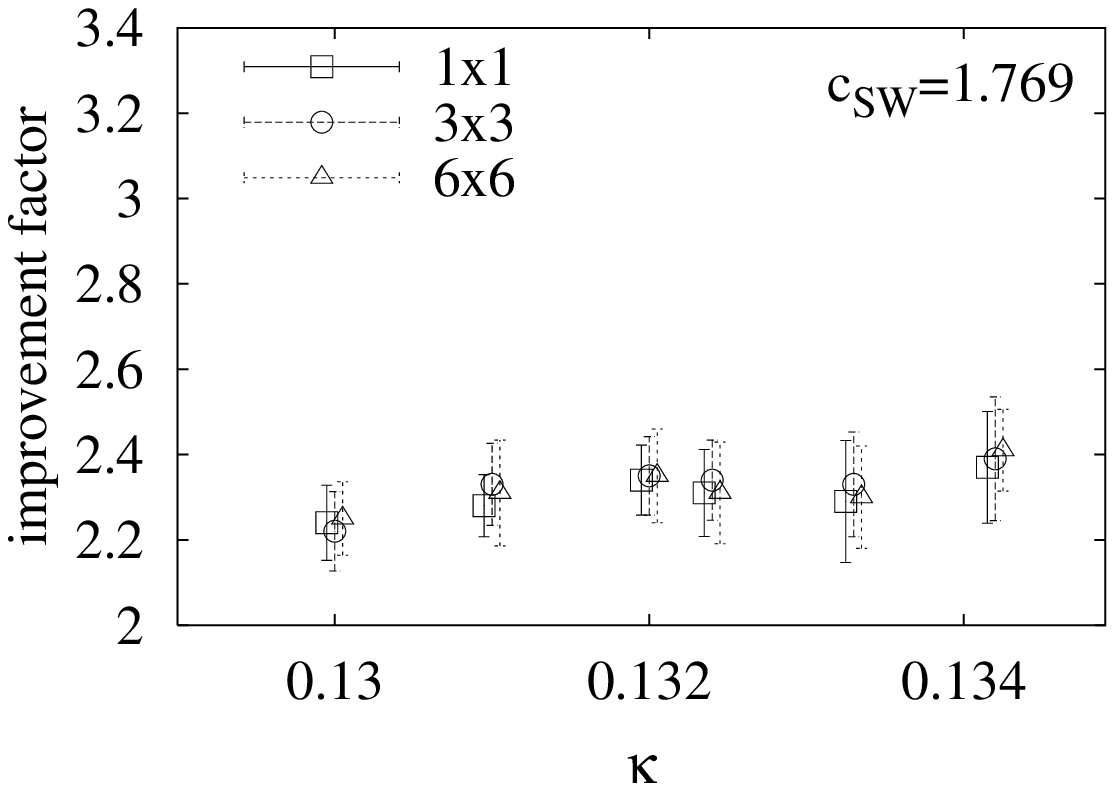}}
\vspace*{-1.2cm}
\caption{
  Gain of \llssor\ over o/e vs. $\kappa$.
\label{OESSOR}\vspace*{-.6cm}}
\end{figure}
We plot the ratio of iteration numbers of the odd-even procedure vs.\ 
\llssor\ as function of $\kappa$ in \fig{OESSOR}. A gain factor up to
2.5 in iteration numbers can be found.  There is no dependence on
$c_{SW}$ or on the block size of $D$ and only 10 \% on the local
lattice size.  As to real CPU costs on APE100, the optimal block size
of $D$ is a $3\times 3$ block whereas on a scalar system, the optimum
is found for a $1\times 1$ diagonal.

Limited by the number of hyper-links to store on the DEC system, we
decided to investigate HF on a lattice of size $8^4$.  We measured at
$\beta=6.0$ in quenched QCD.  We have assessed the critical mass
parameter to determine the critical region of HF. For HF $\omega\simeq
1.0$ is optimal.  We find that SSOR preconditioning of HF leads to
gain factors $\simeq 3$ close to the critical bare mass $m_{c}=-0.92$.

\begin{figure}
{\includegraphics[width=.94\columnwidth]{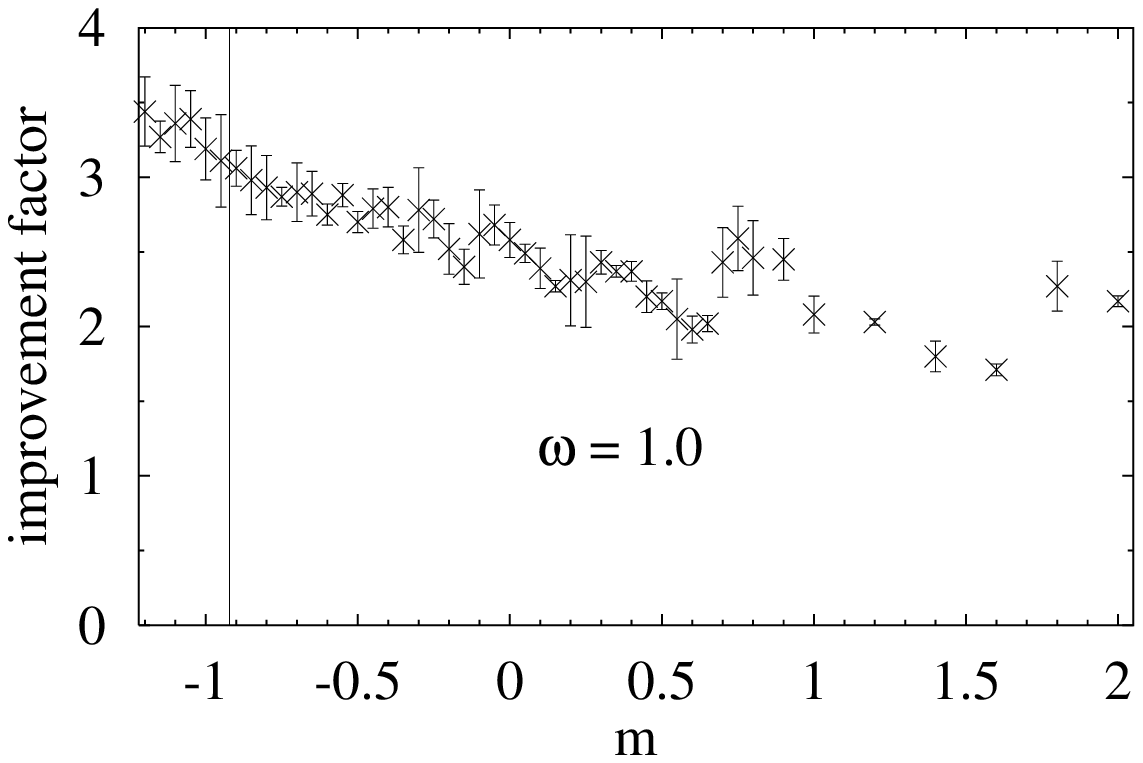}}
\vspace*{-1.4cm}
\caption{
  Dependence of the solution (in CPU time) on the mass parameter $m$.
\label{PERFECT:DEP}\vspace*{-.6cm}}
\end{figure}

\end{document}